\documentclass{appolb}
\usepackage{graphicx}

\usepackage{sidecap}

\usepackage{amsfonts}
\usepackage{amssymb}
\usepackage{amsmath}

\usepackage{dsfont}
\usepackage{cancel}
\usepackage{graphicx}

\usepackage{bm}
\usepackage{epsfig}
\usepackage{psfrag}

\begin{document}
\title{Electroweak hadron structure\\ within a point-form approach%
\thanks{Presented by M. G\'omez-Rocha at Light Cone 2012, Cracow, Poland, 8-13 July, 2012.
Supported by the austrian FWF DK W1203-N16.}%
}
\author{Mar\'ia G\'omez-Rocha, Wolfgang Schweiger
\address{Institut f\"ur Physik, Universit\"at Graz, A-8010 Graz, Austria}
}
\maketitle
\begin{abstract}
We present a relativistic point-form approach for the calculation of
electroweak form factors of few-body bound states. As an example,
the transition form factors for the semileptonic weak decay $B\to
D^*e\bar \nu_e$ are discussed and it is sketched how they can be
extracted unambiguously from the invariant transition amplitude that
describes the process. It is shown how these form factors go over
into one universal function, the Isgur-Wise function in the
heavy-quark limit, $m_Q\to \infty$, and comparison with the
available experimental data is made.
\end{abstract}
\section{The point form of relativistic quantum mechanics}
The point form is one of the three prominent forms proposed by Dirac
in his seminal paper of 1949~\cite{Dirac:1949cp} to formulate
relativistic Hamiltonian dynamics. It  has the nice feature that the
whole Lorentz group (rotations and boosts) is kinematical, i.e. is
not affected by interactions. This allows to boost and rotate
bound-state wave functions in a simple way. As a price, all
components of the 4-momentum operator become interaction dependent.
The formalism presented here is based on the point form of
relativistic quantum mechanics and makes use of the Bakamjian-Thomas
construction~\cite{Bakamjian:1953kh,Keister:1991sb} for introducing
interactions in a fully Poincar\'e invariant manner. As a
consequence the 4-momentum operator factorizes into an interacting
mass operator and a free velocity operator so that it suffices to
consider only an eigenvalue problem for the mass operator. The
formalism presented here has been applied successfully to the study
of electromagnetic properties of spin-0 and spin-1 two-body bound
states consisting of equal-mass
particles~\cite{Biernat:2009my,Biernat:2010tp,Biernat:2011mp}, as
well as to the electroweak structure of mesons consisting of
constituents with different
masses~\cite{GomezRocha:2012zd,Rocha:2010wm,GomezRocha:2011qs}.

The starting point of all these calculations is the physical
processes in which the form factors can be measured, i.e.
electromagnetic scattering or weak decays. To describe these
processes in a fully Poincar\'e invariant manner a multichannel
version of the Bakamjian-Thomas
construction~\cite{Bakamjian:1953kh,Keister:1991sb} is employed. As
mentioned already, the 4-momentum operator then factorizes into an
interacting mass operator and a free 4-velocity operator:
\begin{equation}
 \hat P^\mu= \hat P^\mu_{\text{free}}+\hat P^\mu_{\text{int}}
=\hat M \hat V^\mu_{\text{free}}=(\hat M_{\text{free}}+\hat M_{\text{int}})
\hat V^\mu_{\text{free}}\, .
\end{equation}
The (free) 4-velocity operator $\hat V^\mu_{\text{free}}$ is defined
by $\hat V^\mu_{\text{free}}:=\hat P^\mu_{\text{free}}/\hat
M_{\text{free}} =\hat P^\mu/\hat M$ and describes the overall motion
of the system. The mass operator $\hat M$, depending on internal
variables only, is the quantity of interest, since it contains the
information on the internal structure of the system.

\section{Extracting electroweak currents and form factors}
Electron-meson scattering, e.g., is then formulated on a Hilbert
space consisting of a $e q\bar q$ and a $e q\bar q \gamma$ sector. A
convenient basis consists of, so-called, \textit{velocity states}.
These are multiparticle states characterized by the overall
4-velocity and the center-of-mass momenta and spins of its
components~\cite{Klink:1998zz}. The mass eigenvalue equation to be
solved has the form:
\begin{eqnarray}\label{mass:eq}
 \left(\begin{array}{cc} \hat M_{eq\bar q}^{\text{conf}} & \hat K \\ \hat K^\dagger &
 \hat M_{eq\bar q\gamma}^{\text{conf}}\end{array}\right)
 \left(\begin{array}{c}  |\psi_{eq\bar q}\rangle \\
 |\psi_{eq\bar q\gamma}\rangle \end{array}\right)   =
 m \left(\begin{array}{c} |\psi_{eq\bar q}\rangle \\
 |\psi_{eq\bar q\gamma}\rangle \end{array}\right)\, .
\end{eqnarray}
The diagonal elements of the mass matrix contain the relativistic
kinetic energies and an instantaneous confining interaction between
the quarks. The transition between both channels is caused by $\hat
K^\dagger$ and $\hat K$, which are vertex operators that account for
the creation and annihilation of one photon, respectively. They are
uniquely related to the interaction Lagrangian density of
QED~\cite{Klink:2000pp}. Eliminating the $e q\bar q \gamma$ channel
one ends up with an equation for the $e q\bar q$ component:
\begin{equation}
 (\hat M^{\text{conf}}_{eq\bar q}-m)|\psi_{eq\bar q}\rangle =
\underbrace{\hat K (\hat M^{\text{conf}}_{eq\bar q\gamma}-m)^{-1}
\hat K^\dagger}_{\hat V_{\text{opt}}(m)}|\psi_{eq\bar q}\rangle\, .
\end{equation}
$\hat V_{\text{opt}}(m)$ is an optical potential that describes the
(dynamical) 1-photon exchange between electron and (anti)quark.
On-shell matrix elements of $\hat V_{\text{opt}}(m)$ between
(velocity) states of a confined $q\bar{q}$ pair with quantum numbers
of the meson $M$ provide the invariant 1-photon-exchange amplitude
from which the electromagnetic current of the meson $M$ can be
extracted:
\begin{eqnarray}
 &&\langle V'; \vec k_e',\mu^\prime_e;\vec k_M',\mu^\prime_M|
 \hat V_{\text{opt}}(m)|V;\vec k_e, \mu_e;\vec k_M,\mu_M\rangle_{\text{on-shell}}\nonumber\\
&&\qquad\propto V^0\delta^3(\vec V -\vec V')\frac{j_\mu(\vec
k_e',\mu^\prime_e;\vec k_e,\mu_e)J^\mu(\vec k_M',\mu^\prime_M;\vec
k_M,\mu_M)}{(k_e'-k_e)^2}.
\end{eqnarray}
This relation determines the hadron current and thus the
electromagnetic form factors in a unique way and it fixes also the
normalization of the form factors.
It can be shown~\cite{Biernat:2009my,Biernat:2011mp} that the
resulting electromagnetic current transforms covariantly under
Lorentz transformations and it is conserved for pseudoscalar mesons.
However, because of cluster separability problems inherent in the
Bakamjian-Thomas construction~\cite{Keister:1991sb} one observes
that the current obtained in this way cannot be decomposed in terms
if hadronic covariants only, but one needs additional covariants
built with the sum of the incoming and outgoing electron
4-momenta~\cite{Biernat:2010tp,Biernat:2011mp,GomezRocha:2012zd}.
The need of such additional, spurious covariants resembles the
situation in the covariant front-form
formalism~\cite{Carbonell:1998rj}. There one also encounters
spurious dependencies of the currents on a 4-vector that specifies
the orientation of the light front. Remarkably, if we let the
invariant mass of the electron-meson system go to infinity our
electromagnetic form factors turn out to agree with the front-form
results computed in the $q^+=0$
frame~\cite{Biernat:2009my,Biernat:2010tp,Biernat:2011mp}.\footnote{For
a comprehensive discussion of how to handle cluster problems in
electromagnetic form factors of pseudoscalar and vector bound states
of equal-mass constituents, see Ref.~\cite{Biernat:2011mp}. A more
detailed analysis of cluster problems in heavy-light systems, their
elimination in the heavy quark limit and the connection of point-
and front-form results can be found in
Ref.~\cite{GomezRocha:2012zd}.}

It is quite obvious, how this formalism can be generalized to
semileptonic weak decays in order to calculate transition
amplitudes, currents and decay form
factors~\cite{GomezRocha:2012zd}. Decay processes involve time-like
momentum transfers. Unlike scattering the covariant decomposition of
decay currents does not require the introduction of spurious
covariants, neither in pseudoscalar-to-pseudoscalar nor in
pseudoscalar-to-vector meson decays. Form factors can be extracted
unambiguously in a frame-independent way~\cite{GomezRocha:2012zd}.
Analytical and numerical studies of electromagnetic and weak form
factors of heavy-light systems show that the predictions of
heavy-quark symmetry are respected when one of the constituent
masses goes to infinity~\cite{GomezRocha:2012zd}. As it should be,
one ends up with the, so-called, Isgur-Wise
function~\cite{Isgur:1989vq,Isgur:1989ed}, i.e. one single,
universal, spin-independent form factor that does not depend on the
mass of the heavy quark. Within our approach the Isgur-Wise function
acquires a simple analytical form:
\begin{equation}\label{eq:IWfinal}
\xi(v\cdot v^\prime)= \int\,
\frac{d^3\tilde{k}_{\bar{q}}^\prime}{4\pi}\,
\sqrt{\frac{\omega_{\tilde{k}_{\bar{q}}}}
{\omega_{\tilde{k}^\prime_{\bar{q}}}}}\,\mathcal{S}\,
\psi^\ast\,(\vert \vec{\tilde{k}}_{\bar{q}}^\prime\vert)\,  \psi
 \,(\vert \vec{\tilde{k}}_{\bar{q}}\vert)\, ,
\end{equation}
with $v$ and $v'$ denoting the initial and final meson 4-velocities,
respectively.
$\mathcal{S}=\left(m_{\bar{q}}+\omega_{\tilde{k}^\prime_{\bar{q}}}+\tilde{k}^{
\prime 1}_{\bar{q}}\, \sqrt{\frac{{(v\cdot v^\prime)-1}} {{(v\cdot
v^\prime)+1}}}\right)/\left((m_{\bar{q}}+\omega_{\tilde{k}_{\bar{q}}})
(m_{\bar{q}}+\omega_{\tilde{k}^\prime_{\bar{q}}})\right)^{1/2}$ is a
spin-rotation factor and $\omega_{\tilde{k}_{\bar{q}}}=
\tilde{k}^{\prime 1}_{\bar{q}}\, \sqrt{(v\cdot v^\prime)^2-1}+
\omega_{\tilde{k}^\prime_{\bar{q}}}\, (v\cdot v^\prime)$ with
$\omega_{\tilde{k}_{\bar{q}}^\prime}=\sqrt{m_{\bar
q}^2+\vec{{\tilde{k}}}^{\prime 2}}$.\footnote{A numerical analysis
carried out in Ref.~\cite{GomezRocha:2012zd} reveals that the
spin-rotation factor $\mathcal S$ is by no means negligible, which
emphasizes the need of an appropriate relativistic treatment of the
spin rotation when boosting bound states.}

\section{The $B\to D^* e\bar \nu_e$ transition form factors}
As an example we present numerical results for the transition form
factors of the semileptonic $B\to D^* e\bar \nu_e$ decay.
Calculations are done with a simple harmonic-oscillator wave
function (see Ref.~\cite{GomezRocha:2012zd}) with oscillator
parameteter $a=0.55$~GeV. The most general covariant decomposition
of the weak $B\to D^*$ transition current is usually written in the
form:
\begin{eqnarray}\label{eq:Jppsvdec}
\lefteqn{J^\nu_{B\rightarrow
D^\ast}(\vec{\underline{p}}_{D^\ast}^{\prime},\underline\sigma^\prime_{D^\ast};
\vec{\underline{p}}_B)= \frac{2 i\epsilon^{\nu\mu\rho\sigma}}
{m_B+m_{D^*}}\,\epsilon^*_\mu(\vec{\underline{p}}_{D^\ast}^{\prime},
\underline\sigma^\prime_{D^\ast})\, \underline{p}'_{D^\ast\rho}\,
\underline{p}_{B\sigma} \, V(\underline{q}^2)}\nonumber\\
&&-
2m_{D^*}\,\frac{\epsilon^*(\vec{\underline{p}}_{D^\ast}^{\prime},
\underline\sigma^\prime_{D^\ast}) \cdot
\underline{q}}{\underline{q}^2}\, \underline{q}^\nu\,
A_0(\underline{q}^2) - (m_B+m_{D^*})\,
\epsilon^{*\nu}(\vec{\underline{p}}_{D^\ast}^{\prime},
\underline\sigma^\prime_{D^\ast})\,
A_1(\underline{q}^2)\nonumber\\&& +
\frac{\epsilon^*(\vec{\underline{p}}_{D^\ast}^{\prime},
\underline\sigma^\prime_{D^\ast}) \cdot \underline{q}}{m_B+m_{D^*}}
\,(\underline{p}_B+\underline{p}_{D^\ast}')^\nu\,
A_2(\underline{q}^2)+2 m_{D^*}\,\frac{\epsilon^*
(\vec{\underline{p}}_{D^\ast}^{\prime},
\underline\sigma^\prime_{D^\ast}) \cdot
\underline{q}}{\underline{q}^2}\, \underline{q}^\nu \,
A_3(\underline{q}^2)\, ,\nonumber\\
\end{eqnarray}
where $2 m_{D^\ast} A_3(\underline{q}^2)= (m_B+m_{D^\ast})
A_1(\underline{q}^2) - (m_B-m_{D^\ast}) A_2(\underline{q}^2)$.

As mentioned above, our microscopic expression for the $B\rightarrow
D^\ast$ transition current is not plagued by spurious contributions
and the covarinat decomposition (\ref{eq:Jppsvdec}) can be applied
directly to extract the decay form factors~\cite{GomezRocha:2012zd}.
Introducing the shorthand notation $
J^\nu(\underline\mu^\prime_{D^\ast}) := J^\nu_{B\rightarrow
D^\ast}(\vec{\underline{k}}_{D^\ast}^\prime,
\underline\mu^\prime_{D^\ast};\vec{\underline{k}}_B)$ and adopting
the same kinematics as in Ref.~\cite{GomezRocha:2012zd} one
encounters 10 non-vanishing spin matrix elements of the current,
namely $J^2(0)$, $J^3(0)$, and $J^\mu(\pm 1)$, $\mu=0,1,2,3$. Taking
into account that $J^\mu(1)$ and $J^\mu(-1)$ are related by parity,
one is left with only 6 different matrix elements, 4 of them being
independent. One can see that $A_0$ and $A_2$ enter only $J^0(1)$
and $J^1(1)$. Thus the set $J^2(0)$, $J^3(0)$, $J^0(1)$ and $J^1(1)$
can be used to extract all the $B\rightarrow D^\ast$ decay form
factors.

\begin{figure}[h]
\begin{center}
\includegraphics[width=0.8\textwidth]{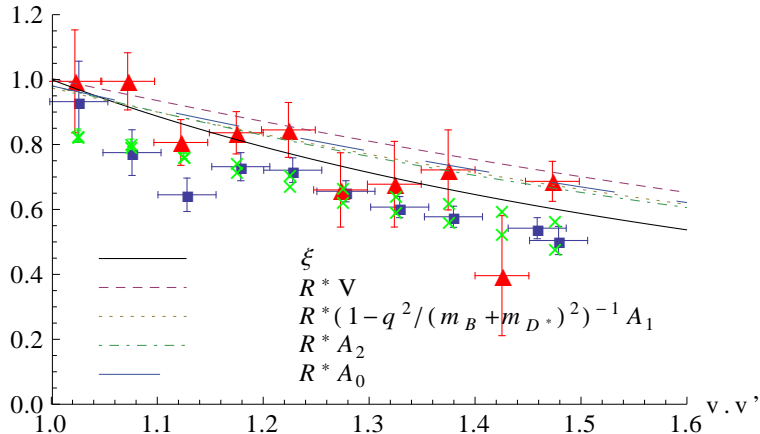}\\
\caption{Weak $B^- \rightarrow D^{0\ast}$ decay form factors
(multiplied by appropriate kinematical factors, $R^*=2\sqrt{m_B
m_D}/(m_B+m_D)$) in comparison with the Isgur-Wise function
$\xi(v\cdot v')$ and with the available experimental data. The
values taken for the physical quark masses are $m_u=0.25$ GeV,
$m_b=4.8$ GeV and $m_c=1.6$ GeV. Experimental data are taken from
Belle~\cite{Abe:2001cs} (dots), CLEO~\cite{Adam:2002uw} (triangles)
and BABAR~\cite{Aubert:2004bw} (crosses) assuming that
$|V_{cb}|=0.0409$, i.e. the central value given by the Particle Data
Group~\cite{PDG2012}.
}
\label{fig:decpsv}
\end{center}
\end{figure}

The numerical results for the form factors as a function of $v\cdot
v'$  (multiplied by appropriate kinematical
factors~\cite{Isgur:1989vq,Isgur:1989ed}) are shown and compared
with the Isgur-Wise function and with the available experimental
data in Fig.~\ref{fig:decpsv}.
Our numerical values for the Isgur-Wise function agree with those
obtained with a front-form quark model~\cite{Cheng:1996if}.
Discrepancies between the point- and front-form approach show up as
soon as the decay form factors are calculated for finite, physical
masses of the heavy quarks. Most likely, these differences can be
attributed to the different roles played by Z-graphs, i.e.
non-valence contributions, in either approach. In the heavy-quark
limit Z-graphs do not contribute to the current, neither in the
front form nor in the point form, which explains why the results
agree for the Isgur-Wise function. For finite quark masses, however,
the inclusion of Z-graphs seems to be crucial for the frame
independence of the decay form factors in front
form~\cite{Cheng:1996if,Bakker:2003up}, whereas this is not the case
in point form (as discussed above).
\vspace{-0.8cm}
\section{Summary and conclusions}
A Poincar\'e invariant description of electromagnetic and weak
currents and form factors of two-body systems has been presented.
The predictions of heavy-quark symmetry are respected by this
approach and a simple analytical expression for the Isgur-Wise
function in terms of the initial and final wave function has been
derived. As an example, numerical results for the semileptonic $B\to
D^* e\bar \nu_e$ decay have been given. Direct comparison of the
Isgur-Wise function with the transition form factors that are
obtained for physical masses of the heavy quarks revealed that
heavy-quark symmetry is broken by about $20\%$. It remains to be
seen, whether Z-graph contributions to the decay process may restore
the equivalence of our point-form approach with front-form
calculations that has already been found for electromagnetic form
factors in the space-like momentum-transfer
region~\cite{Biernat:2009my,Biernat:2010tp,Biernat:2011mp}.


\end{document}